\documentclass[lettersize,journal]{IEEEtran}
\usepackage{amsmath,amsfonts}
\usepackage{bm}
\usepackage{algorithmic}
\usepackage{algorithm}
\usepackage{array}
\usepackage{textcomp}
\usepackage{stfloats}
\usepackage{url}
\usepackage{xurl}
\usepackage{verbatim}
\usepackage{graphicx}
\usepackage{subcaption}
\usepackage{cite}
\usepackage{multirow}
\usepackage{tabularx}
\usepackage{comment}
\usepackage[dvipsnames]{xcolor}
\begin{document}

\renewcommand{\arraystretch}{1.3} 

\title{Any-to-any Speaker Attribute Perturbation for Asynchronous Voice Anonymization}

\author{Liping Chen,~\IEEEmembership{Senior Member,~IEEE}, Chenyang Guo, Rui Wang,\\ Kong Aik Lee,~\IEEEmembership{Senior Member,~IEEE}, and Zhenhua Ling,~\IEEEmembership{Senior Member,~IEEE}

\thanks{Liping Chen, Chenyang Guo, Rui Wang, and Zhenhua Ling are with the University of Science and Technology of China, China (e-mail: lipchen@ustc.edu.cn, chenyangguo@mail.ustc.edu.cn, wangrui256@mail.ustc.edu.cn, zhling@ustc.edu.cn). Kong Aik Lee is with the Hong Kong Polytechnic University, China (e-mail: kong-aik.lee@polyu.edu.hk).}
\thanks{\emph{Corresponding author: Liping Chen}.}
\thanks{This work was supported in part by the Fundamental Research Funds for the Central Universities WK2100000043 and the National Natural Science Foundation of China under Grant U23B2053.}}

\markboth{Journal of \LaTeX\ Class Files,~Vol.~14, No.~8, August~2021}%
{Shell \MakeLowercase{\textit{et al.}}: A Sample Article Using IEEEtran.cls for IEEE Journals}


\maketitle

\begin{abstract}

Speaker attribute perturbation offers a feasible approach to asynchronous voice anonymization by employing adversarially perturbed speech as anonymized output. In order to enhance the identity unlinkability among anonymized utterances from the same original speaker, the targeted attack training strategy is usually applied to anonymize the utterances to a common designated speaker. However, this strategy may violate the privacy of the designated speaker who is an actual speaker. To mitigate this risk, this paper proposes an any-to-any training strategy. It is accomplished by defining a batch mean loss to anonymize the utterances from various speakers within a training mini-batch to a common pseudo-speaker, which is approximated as the average speaker in the mini-batch. Based on this, a speaker-adversarial speech generation model is proposed, incorporating the supervision from both the untargeted attack and the any-to-any strategies. The speaker attribute perturbations are generated and incorporated into the original speech to produce its anonymized version. The effectiveness of the proposed model was justified in asynchronous voice anonymization through experiments conducted on the VoxCeleb datasets. Additional experiments were carried out to explore the potential limitations of speaker-adversarial speech in voice privacy protection. With them, we aim to provide insights for future research on its protective efficacy against black-box speaker extractors \textcolor{black}{and adaptive attacks, as well as} generalization to out-of-domain datasets \textcolor{black}{and stability}. Audio samples and open-source code are published in \url{https://github.com/VoicePrivacy/any-to-any-speaker-attribute-perturbation}.
\end{abstract}

\begin{IEEEkeywords}
Asynchronous voice anonymization, speaker attribute perturbation, speaker-adversarial speech, any-to-any anonymization.
\end{IEEEkeywords}

\section{Introduction}
\IEEEPARstart{D}{riven} by the rapid progress in deep learning, speaker recognition techniques have gone through remarkable progress in recent years. By leveraging speaker recognition techniques\cite{snyder17_interspeech,ECAPATDNN,zeinali2019but,xi-vector}, one can now recognize the identity of a speaker with high accuracy. Besides, advancements in voice conversion (VC) \cite{voice_conversion_overview} and text-to-speech (TTS) \cite{tan2021surveyneuralspeechsynthesis} techniques facilitate the generation of high-quality artificial speech. Consequently, voice privacy has become increasingly susceptible to potential threats. For example, an attacker could retrieve a target speaker's recordings from a breached data source with just a few seconds (e.g., 3 seconds) of their speech, leading to the leakage of privacy-related information such as age, interests, opinions, ethics, and health status. Utilizing VC and TTS techniques, synthetic speech can be generated to impersonate the target speaker. The synthesized speech can be exploited for malicious purposes, such as damaging the speaker's reputation and manipulating public opinion. \textcolor{black}{Given such threats, the need for deepfake speech detection \cite{yamagishi21_asvspoof,safeear} and} voice privacy protection techniques \cite{voiceprivacy} is increasing to prevent the malicious use of speaker information. 

Among privacy protection techniques, voice anonymization, originated in the 1980s \cite{analog_voice_anonymization_TFSP,analog_voice_anonymization_time_frequency}, has regained attention. Currently, voice anonymization can be achieved through either synchronous \cite{Vaidya,SpeechSanitizer,McAdam,voicemask,Hidebehind, 10945923} or asynchronous \cite{v-cloak,chen2023voicecloak,chen2024adversarial,MitigatingTTS,wang2024async} approaches. Specifically, synchronous anonymization modifies both machine and human perceptions, while asynchronous anonymization alters only machine-discernible speaker attribute, maintaining human perception. Among others, facilitated by the speech generation technique, speaker attribute substitution has emerged as the state-of-the-art method for synchronous voice anonymization\cite{fang2019speaker,srivastava2022privacy,asrbn,PSDUtt,nac}. Moreover, in \cite{wang2024async}, the speech generation framework was explored for asynchronous anonymization with limited success.

The adversarial attacks on automatic speaker verification (ASV) \cite{VESTMAN202036,9053076,Fooling2018,SpoofingASV} and identification \cite{zhang2023imperceptible, shamsabadi2021foolhd,9102886, 9053058, 10015819, LAN2022102526,Zhang2022ImperceptibleBW,chen2021real} systems were demonstrated recently. As shown in \cite{v-cloak,chen2023voicecloak,voice_guard,chen2024adversarial,MitigatingTTS}, adversarial perturbation technique provides another viable solution to asynchronous voice anonymization. In these studies, the adversarial perturbation signal was generated by attacking a surrogate speaker extractor using both untargeted and targeted settings. With untargeted-attack setting, the anonymized voice is recognized as a different speaker, thereby achieving de-identification. Nevertheless, untargeted attack training ignores to constrain the speaker similarity among the anonymized utterances. As a result, the anonymized utterances from the same speaker will tend to cluster together, rendering them linkable. The targeted attack setting improves the identity unlinkability capability by anonymizing the original utterances to a shared designated speaker. However, an actual speaker has always been used as the designated speaker, thus violating the voice privacy of the speaker as a result.


\begin{figure*}[!t]
\centering
\includegraphics[scale=1.1]{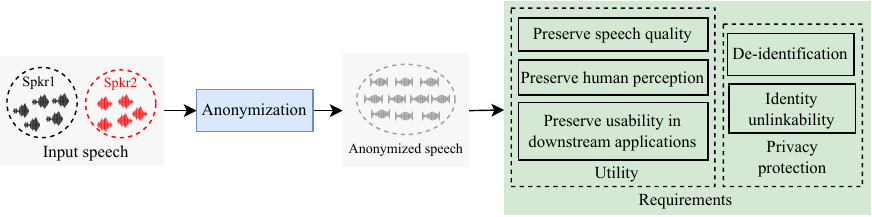}
\caption{Task definition of asynchronous voice anonymization. Input speech utterances from two speakers, \emph{spkr1} and \emph{spkr2}, are represented as black and red dotted ellipses, respectively.}
\label{fig. task_definition}
\end{figure*}

In the context of the untargeted attack setting, to enhance identity unlinkability of anonymized speech while preserving the privacy of actual speakers, \textcolor{black}{this paper investigates an any-to-any training strategy for the adversarial speech generator. It is fulfilled by defining a batch mean loss to supervise its training process.} Specifically, within a training mini-batch, utterances from any speaker are anonymized as a shared pseudo-speaker approximated as the average speaker in the batch. Based on this, a speaker-adversarial speech generation model is proposed in this paper. Existing V-Cloak\cite{v-cloak} and VoiceCloak\cite{chen2023voicecloak} methods generate adversarial speech through transformations of the original version. Different from them, the proposed model employs an adversarial perturbation generator to create speaker attribute perturbations from the original speech, which are then added to the original signal to yield adversarial speech. Specifically, unlike the convolutional neural network (CNN) layers \cite{Simonyan15} used for adversarial transformations in V-Cloak and VoiceCloak, the perturbation generator in our proposed model utilizes the Conformer block \cite{Conformer} as its basic component, leveraging contextual information for adversarial perturbation generation. \textcolor{black}{Furthermore, during the inference process, unlike the VoiceCloak method, which explicitly generates a specific pseudo-speaker, the proposed method does not assign a specific pseudo-speaker identity to the anonymized speech.} Our experiments were conducted on the VoxCeleb dataset, wherein the optimization-based adversarial sample generation methods, MI-FGSM \cite{dong2018boosting} and GRA \cite{zhu2023boosting}, and the feedforward method, V-Cloak, were compared. The results demonstrate the effectiveness of the proposed model in asynchronous voice anonymization. Thereafter, further experiments were carried out on the speaker-adversarial speech generation methods to examine their limitations for voice anonymization, providing insights for future research. The contributions of this paper are summarized as follows:

1) 
We propose the any-to-any strategy to enhance the identity unlinkability in anonymized speech. This is accomplished by introducing a batch mean loss to anonymize the utterances within a training mini-batch to a pseudo-speaker derived as the average speaker within the mini-batch.

2) We propose a speaker-adversarial speech generation model that leverages the any-to-any strategy. It generates anonymized speech by first producing adversarial perturbations and then adding them to the input speech. The effectiveness of the proposed model in asynchronous voice anonymization is validated via experiments conducted on the VoxCeleb dataset.


The rest of this paper is organized as follows. Section \ref{sec. task_definition} defines the task investigated in this paper. The existing speaker-adversarial speech generation methods are briefly described in Section \ref{sec. speaker-adversarial speech}. Our proposed model for any-to-any speaker-adversarial speech generation is presented in Section \ref{sec. any-to-any perturbation}. Section \ref{sec. experiments} presents the experiments and Section \ref{sec. conclusion} concludes this paper.

\section{Task definition}
\label{sec. task_definition}
Fig. \ref{fig. task_definition} depicts the task definition of asynchronous voice anonymization investigated in this work. Given the original speech utterances from different speakers (e.g., spkr1 and spkr2 in Fig. \ref{fig. task_definition}), anonymization is applied before releasing the speech to the public. The requirements for anonymized speech are defined based on the goal of utility and privacy protection. Specifically, the utility requirements include (a) \textbf {preserving speech quality}--preserving the speech quality, (b) \textbf {preserving human perception}--preserving the speaker attribute as perceived by human hearing, (c) \textbf {preserving usability in downstream applications}--maintaining the applicability in downstream tasks. The privacy protection capability requires (d) \textbf{de-identification}--hiding speaker identity from authentic and (e) \textbf{identity unlinkability}--the anonymized utterances from the same original speaker are unlinkable, and those from different original speakers (such as spkr1 and spkr2 in Fig. \ref{fig. task_definition}) are indistinguishable.

\begin{figure}[t]
    \centering
    \begin{subfigure}{0.38\textwidth} 
        \hfill
        \includegraphics[scale=1]{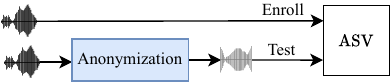}
        \caption{De-identification}
        \label{fig:ignorant}
        \vspace{0.4cm}
    \end{subfigure}
    \hfill
    \begin{subfigure}{0.38\textwidth} 
        \includegraphics[scale=1]{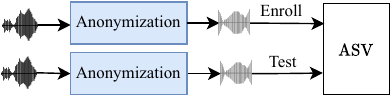}
        \caption{Identity unlinkability}
        \label{fig:lazy-informed}
        \vspace{0.34cm}
    \end{subfigure}
    \caption{ASV evaluation scenarios for de-identification and identity unlinkability.}
    \label{fig:ASV evaluation scenarios}
\end{figure}

As illustrated in Fig. \ref{fig:ASV evaluation scenarios}, the de-identification and identity unlinkability are measured with ASV evaluations as follows:

\noindent $\bullet$ De-identification: As illustrated in Fig. \ref{fig:ASV evaluation scenarios}(\subref{fig:ignorant}), the original speech is used for speaker enrollment while the anonymized speech is used for test. The speaker similarity between the original utterance and its anonymized version is evaluated with the ASV model, measuring \emph{de-identification} capability.

\noindent $\bullet$ Identity unlinkability: As presented in Fig. \ref{fig:ASV evaluation scenarios}(\subref{fig:lazy-informed}), the anonymized speech is used for both enrollment and test. The speaker similarity among the anonymized utterances is evaluated with the ASV model, indicating the \emph{identity unlinkability} capability.

\section{Speaker Attribute Perturbation}
\label{sec. speaker-adversarial speech}
Adversarial perturbation techniques can be categorized into optimization-based and feedforward approaches. This section provides a brief overview of these two methods as well as their applications in speaker attribute perturbation generation.

\subsection{Optimization-based methods}
The fast gradient sign method (FGSM) \cite{FGSM} is the most widely used optimization-based adversarial perturbation technique. In FGSM, given a neural network model and an input sample, a perturbation signal is generated in a single step along the gradient direction maximizing the neural network loss function. Based on that, the iterative FGSM (I-FGSM) \cite{kurakin2017adversarial} is implemented by iterating the FGSM for multiple times. Following that, a momentum strategy was introduced, resulting in the momentum I-FGSM (MI-FGSM). To enhance further the transferability of adversarial perturbations, variants of MI-FGSM have been investigated, such as Nesterov I-FGSM (NI-FGSM) \cite{linnesterov}, NRI-FGSM \cite{NRIFGSM}, and gradient relevance attack (GRA) \cite{zhu2023boosting}.

In the MI-FGSM algorithm, assume a neural network function $\boldsymbol{y}=\mathcal{F}\left( {{\boldsymbol{x}};\theta } \right)$ with model parameters $\theta$. The loss function for adversarial sample generation is defined as $L\left(\boldsymbol{y}\right)$. Given an original sample $\boldsymbol{x}$, the iteration for adversarial sample generation is implemented as follows:
\begin{equation}
    \label{eq. MI-FGSM 1}
    {{{\tilde{\boldsymbol x}}}_{i + 1}} = {\rm{Clip}}^\epsilon \left\{ {{{{\tilde{\boldsymbol x}}}_i} + \alpha \cdot {\rm{sign}}\left( {{{\boldsymbol{g}}_{i + 1}}} \right)} \right\},
\end{equation}
where $i=0,1,2...,I-1$ indicates the iterations with $I$ denoting the number of iterations. The perturbed sample ${{{\tilde{\boldsymbol x}}}_0}$ is initialized to ${\boldsymbol{x}}$. The step size $\alpha$ is constrained within $0<\alpha<\epsilon$. The ${\rm Clip}^{\epsilon}$ function limits $\tilde{\boldsymbol x}$ to the vicinity of $\boldsymbol x$ satisfying the $L_{\infty}$ norm bound with $\epsilon$ representing the intensity. The variable ${\boldsymbol g}_i$ accumulates the gradients from the $i$-th iteration with a momentum decay factor ${\eta}$ as follows:
\begin{equation}
    \label{eq. MI-FGSM 2}
    {{\boldsymbol{g}}_{i + 1}} = \eta \cdot{{\boldsymbol{g}}_i} + \frac{{{\nabla _{\tilde{\boldsymbol{x}}_i}}\left( {L\left( {{\tilde{\boldsymbol{y}}_i} } \right)} \right)}}{{{{\left\| {{\nabla _{\tilde{\boldsymbol{x}}_i}}\left( {L\left( {\tilde{{\boldsymbol{y}}}_i } \right)} \right)} \right\|}_1}}},
\end{equation}
where $\tilde{{\boldsymbol{y}}}_i=\mathcal{F}\left({\tilde{\boldsymbol{x}_i}};\theta\right)$. When $\eta=0$, it reduces to I-FGSM, and to FGSM when $I=0$. After all the $I$ iterations, ${\tilde{\boldsymbol{x}}_I}$ is obtained as the adversarially perturbed sample ${\tilde{\boldsymbol{x}}}$.

In the context of speaker attribute perturbation, the loss function \( L(\boldsymbol{y}) \) can be defined in terms of both untargeted and targeted attacks. Specifically, it can be formulated on the speaker classification loss \cite{shamsabadi2021foolhd,9102886,9053058} or the speaker similarity loss calculated from speaker embedding vectors \cite{10015819,chen2024adversarial}.

\subsection{Feedforward methods} In a feedforward adversarial sample generation framework, adversarial samples are obtained through specific adversarial modules, which are supervised by loss functions \cite{10015819, 9053058, 9102886, shamsabadi2021foolhd,LAN2022102526,odysseyfeedbackVC,chen2021real,v-cloak,chen2023voicecloak,Zhang2022ImperceptibleBW,zhang2023imperceptible,inaudible,illusion,ImpactsOnASR}. The adversarial modules either apply a transformation to the original sample to generate its adversarial version or generate adversarial perturbations that are added to the original sample\cite{9102886,9053058,10015819}. When applied in speaker-adversarial speech generation, the adversarial modules are conventionally supervised by attacking a surrogate speaker extractor. The attack is achieved by either misleading the extractor to identify the speaker within the adversarial speech as a non-original individual (untargeted attack)\cite{chen2021real} or as a designated speaker (targeted attack)\cite{9102886}. Additional supervision can be defined based on the requirements of adversarial speech with respect to its perceptual quality \cite{shamsabadi2021foolhd,inaudible} and performance in downstream tasks \cite{illusion,ImpactsOnASR}. Particularly, in the V-Cloak \cite{v-cloak} and VoiceCloak \cite{chen2023voicecloak} methods, the adversarial utterances were generated via transformations on the original ones, with the transformation supervised by both targeted and untargeted attack loss functions.

In the following, the angular loss \cite{10015819} and perceptual loss \cite{shamsabadi2021foolhd} functions which are used in our work are briefly described.

$\bullet$ \emph{Angular loss:} Given a speaker extractor $\mathcal{F}\left(\bullet\right)$, the speaker attribute within an original utterance $\mathcal{O}$ can be extracted and represented as ${\boldsymbol z}=\mathcal{F}\left(\mathcal{O}\right)$. Assume its speaker-adversarial version ${\tilde{\mathcal O}}$, from which the speaker embedding can be extracted as ${\tilde{\boldsymbol z}}=\mathcal{F}\left(\tilde{\mathcal{O}}\right)$. Mathematically, the angular loss is formulated as:
\begin{equation}
\label{eq. angular loss}
L_{\rm angular}=\frac{{\boldsymbol z}^{\mathsf{T}}\tilde{{\boldsymbol z}}}{{\left\| {\boldsymbol{z}} \right\|_2}{\left\| {\tilde{\boldsymbol{z}}} \right\|_2}},
\end{equation}
where the operator $\left\|  \bullet  \right\|_2$ denotes the L2 norm of the embedding vector. By minimizing the angular loss, the difference between the speaker attributes within the adversarial and original speech is maximized.

$\bullet$ \emph{Perceptual loss:} The perceptual loss \cite
{shamsabadi2021foolhd} was proposed with the aim of preserving the perceptual quality of adversarial speech. It constrains the deviation between the adversarial and original utterances within the acoustic features of their speech frames and is calculated as follows:
\begin{equation}
\label{eq. perceptual loss}
L_{\rm perceptual}=-\frac{1}{T}\sum_{t=1}^{T}\frac{{\boldsymbol f}_t^{\mathsf T}{\boldsymbol {\tilde f}}_t}{{\left\| {\boldsymbol{f}}_t \right\|_2}{\left\| {\boldsymbol{{\tilde f}}}_t \right\|_2}},
\end{equation}
where ${\boldsymbol f}_t$ and $\tilde{{\bf f}_t}$ are the acoustic feature vectors extracted from $\mathcal{O}$ and $\tilde{\mathcal{O}}$, respectively. The variable $T$ is the number of frames.

\begin{figure*}
\centering
\begin{subfigure}{1\textwidth}
\centering
    \includegraphics[scale=0.9]{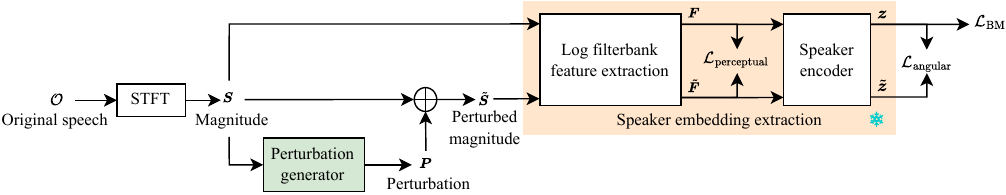}
    \caption{Training flow. The perturbation generator is trainable. The STFT is a signal processing algorithm, and the speaker encoder is pre-trained and frozen during model training.}
    \label{fig:training}
    \vspace{0.4cm}
\end{subfigure}

\hfill

\begin{subfigure}{1\textwidth}
\centering
    \includegraphics[scale=0.9]{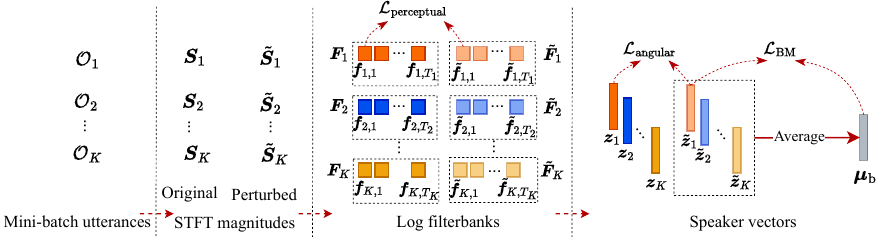}
    \caption{Data flow for training loss computation within a mini-batch of $K$ utterances, i.e., $\left\{{\mathcal O}_1,{\mathcal O}_2,...,{\mathcal O}_K\right\}$. The STFT magnitudes and their perturbed versions are obtained as $\left\{{\boldsymbol{S}_1},{\boldsymbol{S}_2},...,{\boldsymbol{S}_K}\right\}$ and $\left\{\tilde{\boldsymbol{S}}_1,\tilde{\boldsymbol{S}}_2,...,\tilde{\boldsymbol{S}}_K\right\}$. The perceptual loss \( {\mathcal L}_{\rm perceptual} \) is computed between the log filterbank feature vector pairs, i.e., $\left({\boldsymbol{f}}_{k,t},\tilde{\boldsymbol{f}}_{k,t}\right)$ for $k=1,...,K$ and $t=1,...,T_k$. Particularly, the notation $T_k\left(k=1,...,K\right)$ represents the number of frames in ${\mathcal O}_k$. The angular loss \( {\mathcal L}_{\rm angular} \) is calculated between their speaker vector pairs, i.e., $\left({\boldsymbol{z}}_k,\tilde{\boldsymbol{z}}_k\right)$ for $k=1,...,K$. The batch mean loss \( {\mathcal L}_{\rm BM} \) is calculated between the speaker vectors of the adversarial utterances and their average.}
    \label{fig:training_loss_flow}
\end{subfigure}
\hfill

\vspace{0.4cm}

\begin{subfigure}{1\textwidth}
\centering
    \includegraphics[scale=0.9]{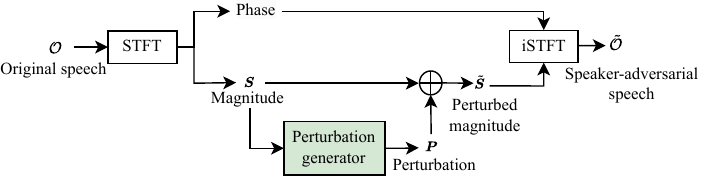}
    \caption{Inference flow. The STFT and iSTFT are signal processing algorithms.}
    \label{fig:inference}
\end{subfigure}

\hfill
\caption{Illustration of any-to-any speaker-adversarial speech generation for asynchronous voice anonymization.}
\label{fig:any-to-any model}
\end{figure*}

\section{Any-to-any Voice Anonymization}
\label{sec. any-to-any perturbation}

\textcolor{black}{Fig. \ref{fig:any-to-any model} illustrates the proposed speaker-adversarial speech generation framework wherein the any-to-any training strategy is applied. The training and inference processes are described in the following.}

\textcolor{black}{\subsection{Training}}

\subsubsection{Overall architecture}
As shown in Fig. \ref{fig:any-to-any model}(\subref{fig:training}), given an input utterance ${\mathcal O}$, the short-time Fourier transform (STFT) is applied to extract the magnitude spectrum ${\boldsymbol S}=\left[{\boldsymbol s}_1,...,{\boldsymbol s}_T\right]$ with $T$ frames. After going through a perturbation generation module, the $T$ perturbation vectors are obtained as ${\boldsymbol P}$. Adding ${\boldsymbol P}$ to ${\boldsymbol S}$ gives the perturbed magnitude vectors as ${\boldsymbol {\tilde S}}=\left[{\tilde{\boldsymbol s}}_1,...,{\tilde{\boldsymbol s}}_T\right]$. Upon ${\boldsymbol S}$ and ${\tilde {\boldsymbol S}}$, the log filterbank feature vectors are extracted as ${\boldsymbol F}$ and ${\boldsymbol{\tilde {F}}}$. Thereafter, a pre-trained speaker encoder is employed to extract the speaker embedding vectors for both the original and perturbed versions of the input utterance, represented as ${\boldsymbol z}$ and ${\boldsymbol {\tilde z}}$, respectively.

In this framework, the perturbation generator is composed of Conformer blocks. The speaker encoder can be pre-trained in various neural speaker classification architectures, such as x-vector\cite{snyder17_interspeech}, ECAPA-TDNN\cite{ECAPATDNN}, ResNet\cite{zeinali2019but}, and xi-vector\cite{xi-vector}. Moreover, it is frozen in the proposed speaker-adversarial speech generation framework.

\subsubsection{Loss function}
As illustrated in Fig. \ref{fig:any-to-any model}(\subref{fig:training_loss_flow}), the training process of the proposed any-to-any speaker-adversarial speech generation model is conducted within mini-batches. Given a mini-batch composed of $K$ utterances $\left\{{\mathcal O}_1,{\mathcal O}_2,...,{\mathcal O}_K\right\}$, the STFT magnitudes are extracted from each of them, represented as $\left\{\boldsymbol{S}_1,\boldsymbol{S}_2,...,\boldsymbol{S}_K\right\}$. Thereafter, their perturbed versions are obtained as $\left\{\tilde{\boldsymbol{S}}_1,\tilde{\boldsymbol{S}}_2,...,\tilde{\boldsymbol{S}}_K\right\}$. Following that, the log filterbank features can be extracted from both the original and perturbed STFT magnitude vectors of the $K$ utterances, denoted as $\left\{\boldsymbol{F}_1,\boldsymbol{F}_2, ...,\boldsymbol{F}_K\right\}$ and $\left\{\tilde{\boldsymbol{F}}_1,\tilde{\boldsymbol{F}}_2, ...,\tilde{\boldsymbol{F}}_K\right\}$, respectively. Following that, the speaker vectors are extracted from them and represented as $\left\{\boldsymbol{z}_1,\boldsymbol{z}_2, ...,\boldsymbol{z}_K\right\}$ and $\left\{\tilde{\boldsymbol{z}}_1,{\tilde{\boldsymbol{z}}}_2, ...,\tilde{\boldsymbol{z}}_K\right\}$.

In our work, a batch mean loss is defined on the training utterances in a mini-batch. Firstly, the average of the speaker vectors extracted from the adversarial utterances is computed as follows:

\begin{equation}
    \label{eq. batch mean mu}
    {\bm \mu}_{\rm b}=\frac{1}{K}\sum_{k=1}^{K}\tilde{{\boldsymbol z}}_k,
\end{equation}
where the subscript $_{\rm b}$ denotes the training batch. The vector ${\bm \mu}_{\rm b}$ represents the average speaker within the mini-batch and serves as the pseudo-speaker for the utterances in the mini-batch. Thereby, the batch mean loss is defined as the cosine distance between the speaker vector of each utterance and the average as follows:
\begin{equation}
{\mathcal L}_{\rm BM}=-\frac{1}{K}\sum_{k=1}^K\frac{\tilde{\boldsymbol z}_k^{\mathsf T}{\bm \mu}_{\rm b}}{{\left\| {{\tilde{\boldsymbol z}}}_k \right\|_2}{\left\| {{{\bm {\mu }}_{\rm{b}}}} \right\|_2}},
\end{equation}
where the subscript $_{\rm BM}$ stands for \emph{batch mean}. By minimizing \( {\mathcal L}_{\rm BM} \), the speaker extractor perceives the speaker attributes of adversarial utterances from different speakers in the training batch as the same pseudo-speaker, represented by \( \bm{\mu}_{\rm b} \).

Under the supervision of the batch mean loss, defined for the anonymized utterances, the linkability of anonymized utterances from the same original speaker is reduced. As well, the distinguishability among those from different original speakers gets reduced. Moreover, as distinct pseudo-speakers are obtained in different mini-batches, the batch mean loss performs an any-to-any anonymization from the original speaker to the pseudo-speaker.


Additionally, the angular loss is calculated from the speaker vectors derived from the original and adversarial utterances to conceal the speaker attribute inherent in the original utterance within its adversarial version, computed as follows:
\begin{equation}
    \label{eq. any-to-any angular loss}
    {\mathcal L}_{\rm angular}=\frac{1}{K}\sum_{k=1}^{K}\frac{{\boldsymbol{z}}^{\mathsf{T}}_k{{\tilde{\boldsymbol{z}}}_k}}{{\left\|{\boldsymbol{z}}^{\mathsf{T}}_k\right\|}_2{{\left\|{\tilde{\boldsymbol{z}}}_k\right\|}_2}}.
\end{equation}

\noindent Furthermore, the perceptual loss is calculated on the log filterbank feature vectors of the original and adversarial utterances to maintain the perceptual quality, defined as follows:
\begin{equation}
    \label{eq. any-to-any perceptual loss}
    {\mathcal L}_{\rm perceptual}=-\frac{1}{\sum_{k=1}^KT_{k}}\sum_{k=1}^K\sum_{t=1}^{T_k}\frac{{\boldsymbol{f}_{k,t}^{\mathsf T}}{\tilde{\boldsymbol{f}}_{k,t}}}{\left\|{\boldsymbol{f}_{k,t}}\right\|_2\left\|{\tilde{\boldsymbol{f}}_{k,t}}\right\|_2}.
\end{equation}
In (\ref{eq. any-to-any perceptual loss}), $\left\{\boldsymbol{f}_{k,1},...,{\boldsymbol{f}_{k,T_k}}\right\}$ and $\left\{\tilde{\boldsymbol{f}}_{k,1},...,{\tilde{\boldsymbol{f}}_{k,T_k}}\right\}$ are the log filterbank feature vectors extracted from the $k$-th original utterance and its adversarial version, respectively, with $T_k$ denoting the number of frames.

Finally, the loss function used in the adversarial speech generation framework is defined as:
\begin{equation}
\label{eq. loss func}
{\mathcal L}={\alpha}{\mathcal L}_{\rm perceptual}+{\beta}{\mathcal L}_{\rm angular}+{\gamma}{\mathcal L}_{\rm BM},
\end{equation}
where $\alpha$, $\beta$, and $\gamma$ are the weights of the three loss functions and $\alpha+\beta+\gamma=1$.

\textcolor{black}{It is noteworthy that with any-to-any training strategy fulfilled by the batch mean loss, the pseudo-speaker identity in the adversarial speech is obfuscated. This ensures that the adversarial speech utterances obtained by introducing the adversarial perturbations generated by the perturbation generator are no longer linked by their original speaker identities.}

\subsection{Inference}
As illustrated in \textcolor{black}{Fig. \ref{fig:any-to-any model}(\subref{fig:inference})}, during inference, given an original speech utterance ${\mathcal{O}}$, the STFT magnitude and phase spectra are extracted first. The perturbation vectors $\boldsymbol{P}$ are then generated based on the magnitude vectors and then added to ${\boldsymbol{S}}$, yielding the perturbed magnitude $\tilde{\boldsymbol{S}}$. Finally, the perturbed speech $\tilde{{\mathcal{O}}}$ is generated with the inverse STFT (iSTFT) using $\tilde{\boldsymbol{S}}$ and the original phase, obtaining the anonymized speech. \textcolor{black}{In the anonymization process of an original speech, no specific pseudo-speaker identity is assigned to the anonymized output, thereby facilitating an any-to-any anonymization.}

\section{Experiments}
\label{sec. experiments}


\subsection{Datasets \& configurations}
Our experiments were carried out on the datasets of VoxCeleb1 \cite{nagrani17_interspeech}, VoxCeleb2 \cite{chung18b_interspeech}, LibriSpeech \cite{LibriSpeech}, and AIShell\cite{aishell_2017}. Among them, the VoxCeleb1\&2 datasets were used for model training. Specifically, for each speaker in VoxCeleb1, a random selection of utterances, ranging from 6 to 11, was preserved as a closed speaker development set. The rest utterances of Voxceleb1 were used for model training together with Voxceleb2 in our experiments, including 2,680 hours of speech from 7,205 speakers. The VoxCeleb1-O test set and the development and test subsets from LibriSpeech and AIShell were used for evaluations.

In our experiments, an ECAPA-TDNN was pre-trained on VoxCeleb1\&2, with its encoder utilized as the speaker extractor for speaker-adversarial speech generation. The training recipe as provided in the open-source toolkit {ASV}-SUBTOOLS\footnote{\url{https://github.com/Snowdar/asv-subtools/blob/master/pytorch/launcher/runEcapaXvector_online.py}} \cite{9414676} was used. In detail, data augmentation was implemented using RIRs\footnote{\url{https://www.openslr.org/28/}} and MUSAN \cite{MUSAN} datasets. The acoustic feature of 40-dimensional log filterbank with a 10ms frameshift was applied. Three SE-Res2Blocks and attentive statistics pooling mechanism \cite{attentive_pooling} were used. The dimension of the speaker embedding was 192.

In the implementation of the proposed speaker-adversarial speech generation method, 256 frequency bins were extracted with the STFT algorithm. The speaker encoder from the pre-trained ECAPA-TDNN model is utilized. The perturbation generator was composed of 6 Conformer blocks. In each block, the convolutional kernel size was 31, and 4 self-attention heads were applied. The sizes of the hidden and the output layers were 1,024 and 256. Rectified linear units (ReLU) \cite{nair2010rectified} was used as the activation function. During training, a warm-up strategy was employed for the learning rate, reaching a peak of 0.001 over 9,600 warm-up steps, followed by an exponential decay. The weight $\alpha$ of the perceptual loss ${\mathcal L}_{\rm perceptual}$ in the loss function equation (\ref{eq. loss func}) was fixed as 0.5. The weight variables $\beta$ and $\gamma$ regarding the angular and batch mean losses were finetuned on the preserved closed dataset from VoxCeleb1, achieving a balance between de-identification and identity unlinkability capabilities. Finally, the weight configuration $\left\{\alpha=0.5,\beta=0.15,\gamma=0.35\right\}$ was adopted.


\subsection{Baseline methods}

In examining the effectiveness of the proposed method in voice anonymization, four methods were experimented with for comparison, including {MI-FGSM}, {GRA}, {V-Cloak}, and the {proposed model without BM loss}. Among these, the MI-FGSM and GRA algorithms were used as representatives of optimization-based speaker-adversarial speech generation methods. V-Cloak served as the baseline speaker-adversarial speech generation technique in asynchronous voice anonymization. Besides, it operates in the feedforward manner, which is the same as our proposed model. The experiments conducted on the w/o BM configuration of our proposed model served as an ablation study to assess the effectiveness of the proposed any-to-any strategy. The detailed configurations are as follows:

$\bullet$ \emph{MI-FGSM \& GRA} -- In our experiments, MI-FGSM was applied as a baseline method, providing a foundational approach for generating adversarial samples that utilized iterative and momentum strategies. GRA presented an enhanced variant of MI-FGSM with improved transferability from white-box to black-box speaker extractors. The loss function was defined as the cosine distance calculated between speaker embedding vectors extracted from the original and adversarial utterances using the ECAPA-TDNN speaker encoder, formulated as follows:
\begin{equation}
\label{eq. optimization-based loss function}
    L\left(\bullet\right)=-\frac{{\boldsymbol z}^{\mathsf{T}}\tilde{{\boldsymbol z}}}{{\left\| {\boldsymbol{z}} \right\|_2}{\left\| {\tilde{\boldsymbol{z}}} \right\|_2}},
\end{equation}
with ${\boldsymbol z}$ and $\tilde{{\boldsymbol z}}$ denoting the respective speaker embedding vectors. In both the MI-FGSM and GRA methods, the adversarial speech was generated with 10 iterations. The attack intensity $\epsilon$ and step size $\alpha$ were set to 0.0012 and 0.00012, respectively. Particularly, the momentum in the MI-FGSM implementation was set as $\eta$=1.4 and the number of nearby samples in the GRA method was 10.

$\bullet$ \emph{V-Cloak} -- The experiments in V-Cloak were conducted following the open-source recipe which is available at \footnote{\url{https://github.com/V-Cloak/V-Cloak}}. \textcolor{black}{Within the model, a VP-Modulation [2] module deriving from the Wave-U-Net structure\cite{Wave_u_net} was employed.} The anonymization model was trained with the VoxCeleb1\&2 datasets and in the untargeted attack manner. To achieve optimal voice privacy protection capability and ensure a fair comparison with our proposed method, the weight $\alpha$ associated with the intelligibility component in the loss function, as outlined in equation (2) of \cite{v-cloak}, was set to 0. Following the strategy presented in \cite{v-cloak}, the rest loss weight parameters $\left\{\beta,\gamma\right\}$ in equation (2) of \cite{v-cloak} for V-Cloak model training were set in a stepwise manner as follows. When $L_{\rm ASV} \geq 0.3$, $\beta$ = 5e-8, $\gamma$ = 0.4. When $0.15 \leq
L_{\rm ASV} < 0.3$, $\beta$ = 1e-7, $\gamma$ = 0.4. When $L_{\rm ASV} < 0.15$, $\beta$ = 1.5e-7, $\gamma$ = 0.4.

$\bullet$ \emph{w/o BM} -- Additionally, an ablation study was conducted in the proposed method by removing the batch mean loss from the loss function, achieved by setting $\left\{\alpha=0.5, \beta=0.5,\gamma=0\right\}$ in equation (\ref{eq. loss func}). In this implementation, voice privacy protection was attained using an untargeted attack loss function, analogous to our experiments on MI-FGSM, GRA, and V-Cloak.

It is noteworthy that, \textcolor{black}{in all compared methods, the speaker encoder from our pre-trained ECAPA-TDNN model was used as the surrogate speaker extractor. Among the compared methods, both the V-Cloak and our proposed method involved a specific adversarial speech generation module. The parameter sizes of the generators in the V-Cloak and our proposed models were 66.90 MB and 34.85 MB, respectively.} The parameters were tuned on the preserved closed dataset from VoxCeleb1 in these methods. In the following, the comparison between the proposed and the four baseline methods is presented regarding their influence on speech quality, human perception, and intelligibility. After that, their capability of voice privacy protection is demonstrated, in terms of de-identification and identity unlinkability. At last, additional experiments are presented to discuss the potential limitations in the application of speaker-adversarial speech for voice privacy protection.

\begin{figure}[!t]
\centering
\includegraphics[scale=0.5]{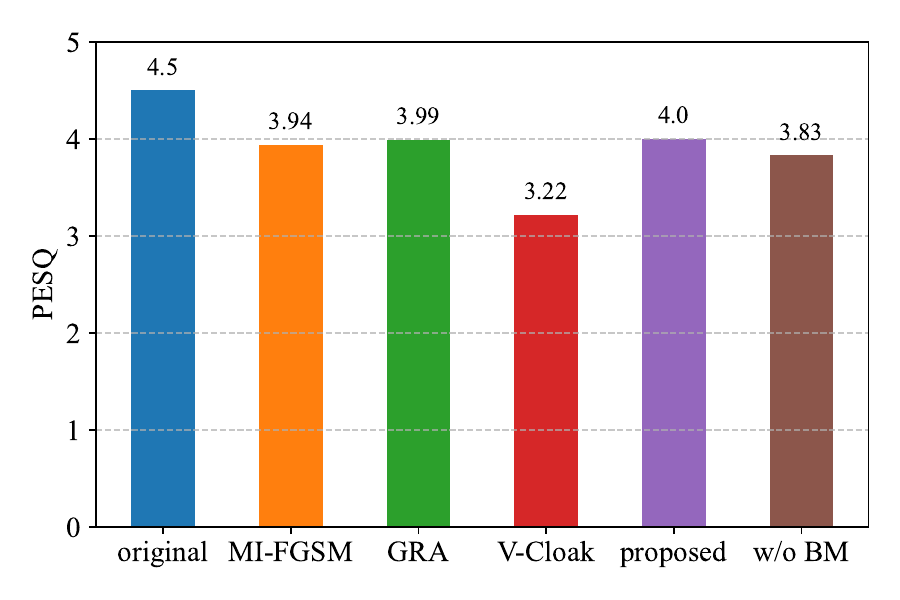}
\caption{PESQ values of the speaker-adversarial speech generated using MI-FGSM, GRA, V-Cloak, and the proposed method, as well as the configuration w/o BM.}
\label{fig:PESQ}
\end{figure}

\subsection{Speech quality assessment}
The speech quality was measured with PESQ, with values ranging from 0 to 4.5. The PESQ values computed on the adversarial utterances obtained by the five baseline methods are presented in Fig. \ref{fig:PESQ}. From the results, it can be observed that our proposed method achieved a PESQ value higher than 4.0, suggesting that the introduction of the speaker attribute perturbation did not induce significant deterioration in speech perceptual quality. Moreover, the proposed method attained the highest PESQ value among all compared approaches. This demonstrates the superiority of the proposed model in preserving speech quality.


\begin{figure}[!t]
\centering
\includegraphics[scale=0.5]{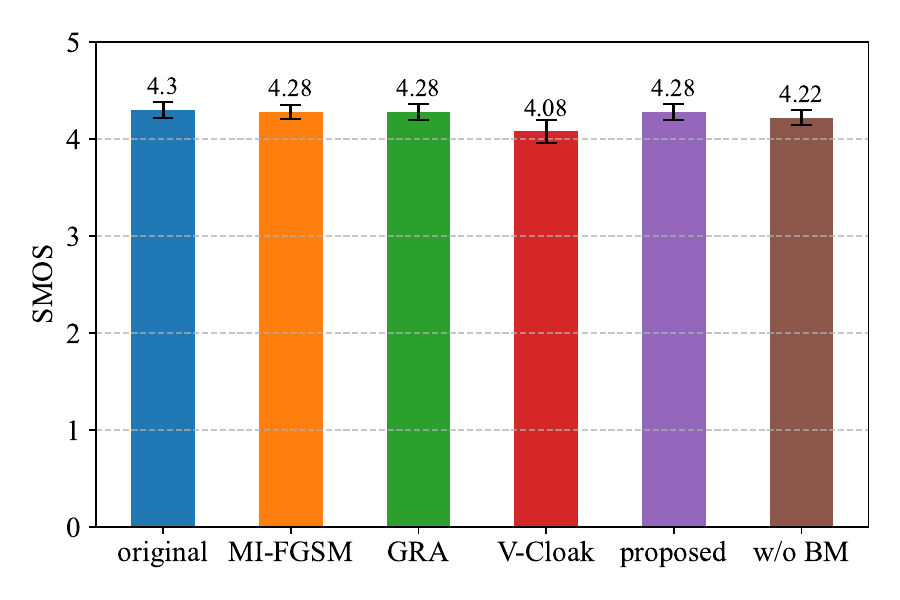}
\caption{The SMOS results between the original recording and the adversarial speech obtained with MI-FGSM, GRA, V-Cloak, and our proposed method (w/o BM included). Additionally, the result obtained between the original recording and itself is included for reference. The results are displayed as original-* test pairs, where * represents the type of utterance analyzed, either the original recording or the adversarial speech generated by the compared methods. The compared methods are indicated on the x-axis.}
\label{fig:SMOS}
\end{figure}

\subsection{Human perceptual evaluation}
Next, a SMOS test was conducted between the speaker-adversarial speech utterances and the original recordings to examine the compared methods in speaker perception preservation. In these tests, five female and five male speakers were randomly selected from the VoxCeleb1-O test utterances, with one utterance randomly chosen from each speaker. The results were collected from 15 paid native speakers who were required to score speaker similarities between speech pairs from 1 to 5, with 0.5 being the step. Fig. \ref{fig:SMOS} presents the SMOS with 95\% obtained with the adversarial speech generated with the MI-FGSM, GRA, V-Cloak, and our proposed methods, including the w/o BM configuration.


From Fig. \ref{fig:SMOS}, it can be observed that the speaker-adversarial speech generated with the MI-FGSM, GRA, and our proposed methods achieved a comparable SMOS score with the original recordings, i.e., 4.28 vs 4.30. This indicates that these methods are capable of preserving the speaker attribute within the original speech when perceived by human hearing. In the proposed method, the omission of batch mean loss resulted in a minor degradation in SMOS, yet yielded a higher SMOS score than the V-Cloak method, which is consistent with the observations regarding PESQ values as shown in Fig. \ref{fig:PESQ}.

\subsection{Speech intelligibility assessment}
\begin{figure}[t]
    \centering
    \includegraphics[scale=0.5]{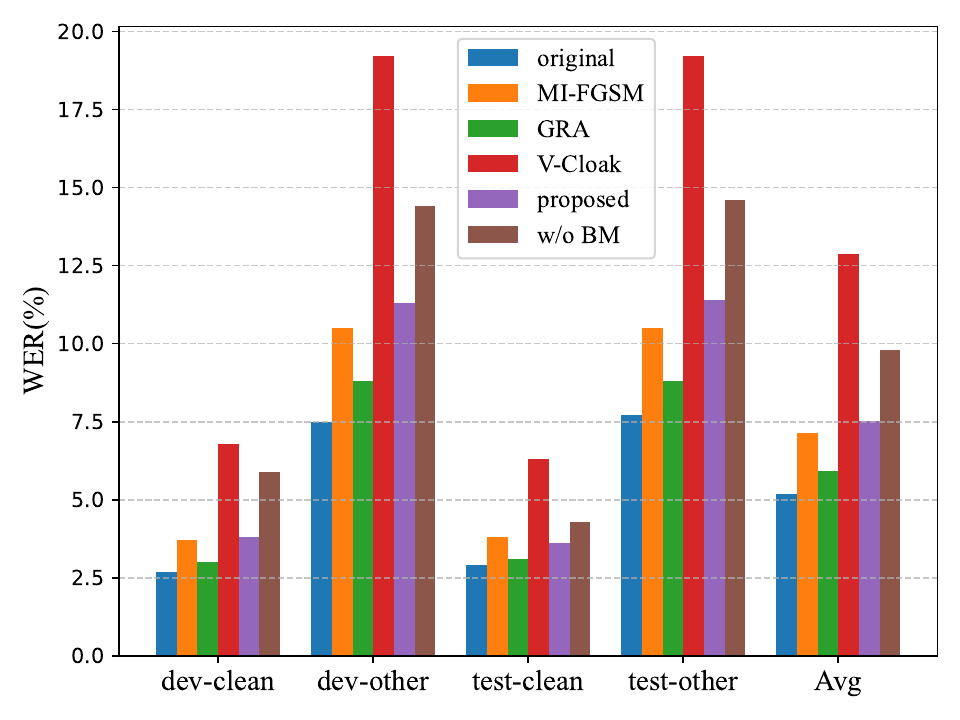}
    \caption{WERs (\%) on the original recording and speaker-adversarial speech on the test-clean, test-other, dev-clean and dev-other sets of LibriSpeech. The original and adversarial utterances generated with MI-FGSM, GRA, V-Cloak, and our proposed method, including the w/o BM configuration, are presented. The average WERs are obtained for each method across the test datasets and presented in the section \emph{Avg}.}
    \label{fig:WER res}
\end{figure}

Speech intelligibility was selected as the indicator of the effect of speaker-adversarial speech on downstream applications, demonstrating their applicability in automatic speaker recognition (ASR). The evaluations were conducted on the test-other, test-clean, dev-clean, and dev-other datasets of LibriSpeech. A Conformer-Transformer ASR model, following the architecture utilized in \cite{Conformer}, was trained on the LibriSpeech libri-train-960 dataset using the open-source ESPnet toolkit\footnote{\url{https://github.com/espnet/espnet}}\cite{hayashi2020espnet}. In the training process, the utterances were augmented with signal-to-noise ratios of range [25, 30]. The word error rates (WERs) obtained by the speaker-adversarial speech generated with the five compared methods are presented in Fig. \ref{fig:WER res}.

As shown in Fig. \ref{fig:WER res}, compared with the original utterances, the introduction of speaker attribute perturbations in the five compared methods led to higher WERs. This indicates that the changes made to the speaker attribute also affect the linguistic information conveyed in the speech as perceived by the ASR model. Specifically, the proposed method achieved comparable WERs to optimization-based approaches, i.e. MI-FGSM and GRA. This demonstrates that its impact on the application of anonymized speech in downstream ASR remains within a manageable range.

\subsection{Voice privacy protection evaluation}
The voice privacy protection capability was measured in terms of de-identification and identity unlinkability, respectively. The evaluations were conducted on the VoxCeleb1-O trials and in the white-box condition. The speaker extractor of the ECAPA-EDNN model, which was used in adversarial speech generation was used for speaker embedding extraction. The cosine similarity computed on the speaker embedding was used as the score. The performances were measured with equal error rates (EERs). Notably, the speaker extractor attained an EER of 1.46\% on the VoxCeleb1-O trials on the original speech utterances. In the de-identification evaluation, the original speech was used for speaker enrollment and the adversarial speech was used for the test. In the identity unlinkability evaluation, adversarial speech was used for both enrollment and test. In these two evaluations, higher EERs indicate better capability of de-identification and identity unlinkability. The EERs are presented in Table \ref{tb. ASV eval res}.

\begin{table}[t]
\caption{EERs(\%) obtained by the speaker-adversarial speech in both de-identification (de-id) and identity unlinkability (id-unlnk) evaluations. The MI-FGSM, GRA, V-Cloak, and our proposed method, including w/O BM, are presented. For reference, the EER on the original speech was $1.46\%$.}
\label{tb. ASV eval res}
\centering
\begin{tabular}{c|c|c|c|c|c}
\hline
&MI-FGSM & GRA & V-Cloak & proposed & w/o BM \\
\hline
de-id & 24.93& 46.44 &36.97 & \textbf{46.79} & 71.48\\
\hline
id-unlnk & 35.71 & 33.11 & 22.39 & \textbf{38.93} & 21.81\\
\hline
\end{tabular}
\end{table}

As shown in Table \ref{tb. ASV eval res}, in the proposed method, when compared with the w/o BM configuration, the absence of the batch mean loss resulted in a higher EER in the de-identification evaluation and a lower EER in the unlinkability evaluation. This indicates that in the proposed method, the proposed any-to-any strategy was effective in enhancing identity unlinkability at the expense of the de-identification capability. However, in the de-identification evaluation, it achieved an EER of 46.79\%, which is close to 50\%, demonstrating its sufficient capability to hide the original speaker's identity in the anonymized version. Such observations indicate that the introduction of the any-to-any strategy enabled our proposed model to attain a balance between de-identification and identity unlinkability capabilities.

Moreover, compared to the MI-FGSM, GRA, and V-Cloak methods, the speaker-adversarial speech generated by our proposed approach without the BM loss function (w/o BM) exhibited superior performance in de-identification. This demonstrates that under untargeted attack training supervision, the proposed model architecture surpassed the baseline methods in hiding identities within speech utterances. Moreover, with the application of the batch mean loss in our proposed method, it achieved the highest EERs in both de-identification and identity unlinkability evaluations compared to the MI-FGSM, GRA, and V-Cloak methods, demonstrating its effectiveness in voice privacy protection.

\subsection{Summary and discussions}

Combining the comparisons conducted on speech quality, human perception, and intelligibility as presented in Figs. \ref{fig:PESQ}-\ref{fig:WER res} and voice privacy protection as presented in Table \ref{tb. ASV eval res}, the performance of the proposed method in asynchronous voice anonymization can be summarized as follows:

1) In our proposed speaker-adversarial speech generation framework, the introduction of the any-to-any strategy, which is fulfilled by the batch mean loss, effectively enhanced identity unlinkability among the anonymized utterances. Moreover, it achieved a balance between the de-identification and unlinkability in the anonymized speech.

2) Compared with the optimization-based MI-FGSM and GRA methods, the proposed method achieved improved de-identification and identity unlinkability, without sacrificing its preservation of the original speech quality, human perception, and intelligibility.

3) In the feedforward speaker-adversarial speech generation methodology, our proposed generation model demonstrated improvements over the V-Cloak method in speech quality, human perception, de-identification, and identity unlinkability.

In the end, additional experiments were conducted to further study the application of the proposed speaker-adversarial speech generation method in voice privacy protection. The transferability of the voice-protection capability was first studied against black-box speaker extractors. Following that, the generalization capability was examined on out-of-domain evaluation datasets.

\subsubsection{Black-box speaker extractors}
\textcolor{black}{The voice privacy protection capability of the speaker-adversarial speech was evaluated with four black-box speaker extractors, including the speaker encoders derived from a ResNet34\cite{zeinali2019but}, an xi-vector\cite{xi-vector}, a CAM++ \cite{CAM++}, and a ResNet100 \cite{SpkrVeriResNet} models.} Specifically, the xi-vector model employed the same speaker encoder structure as our ECAPA-TDNN model. The ResNet34 and xi-vector models were trained with the open-source recipes provided in ASV-SUBTOOLS \footnote{\url{https://github.com/Snowdar/asv-subtools/blob/master/pytorch/launcher/runResnetXvector_online.py}} \footnote{\url{https://github.com/Snowdar/asv-subtools/blob/master/pytorch/launcher/runSnowdar_Xivector.py}}, using the VoxCeleb1\&2 datasets. In the training processes of both models, the data augmentations offered in the scripts were applied, using RIRs and MUSAN. \textcolor{black}{Besides, the CAM++ speaker extractor was derived from the open-source model available at \footnote{https://github.com/lovemefan/campplus/tree/main}, and trained on the VoxCeleb2 dataset. The ResNet100 speaker extractor was derived from the open-source model available at \footnote{https://drive.google.com/drive/folders/1Tl5qJ2Rbk00LdQ7m5HEjKqZi35 P6U3kF}, which was trained with the VoxBlink2 dataset \cite{voxblink2}.} In the ASV evaluations on these four black-box speaker extractors, the cosine distances computed on the speaker embedding vectors were used as the scores. The evaluations were performed on the VoxCeleb1-O test trials. The results are presented in Table \ref{tb. black-box ASV eval res}. Particularly, Table \ref{tb. black-box ASV eval res}(\subref{tb: black-box original}) gives the EERs obtained from original recordings using the ECAPA-TDNN, ResNet34, xi-vector, \textcolor{black}{CAM++, and ResNet100} extractors, serving as a reference for their capabilities in speaker attribute extraction.

\begin{table}[t]
\caption{Evaluation results of the voice privacy protection against black-box speaker extractors.}
\label{tb. black-box ASV eval res}
\centering
    \begin{subtable}[t]{\linewidth}
    \caption{EERs (\%) obtained by the original recordings on the speaker extractors within our ECAPA-TDNN (ECAPA), ResNet34, and xi-vector models.}
    \label{tb: black-box original}
    \centering
    \begin{tabular}{c|c|c|c|c}
    \hline
    ECAPA &ResNet34 & xi-vector & \textcolor{black}{CAM++} & \textcolor{black}{ResNet100 }\\
    \hline
    1.46 & 1.81 &  2.07 & \textcolor{black}{6.17}& \textcolor{black}{0.58}  \\
    \hline
    
    \end{tabular}
    \end{subtable}

    \vspace{0.4cm}

    \begin{subtable}[t]{\linewidth}
    \setlength{\tabcolsep}{1.5mm}
    \caption{EERs(\%) obtained by the speaker-adversarial speech on the black-box speaker extractors of the ResNet34 and xi-vector. The results obtained in the white-box ECAPA-TDNN (ECAPA) extractor are presented for reference. Both the de-identification (de-id) and identity unlinkability (id-unlnk) evaluations are included. The MI-FGSM, GRA, V-Cloak, and our proposed methods, are compared. The column \emph{w/b} indicates white-/black-box extractors.}
    \label{tb: black-box asv res}
    \centering
    \begin{tabular}{c|c|c|c|c|c}
    \hline
    &\multirow{2}{*}{w/b}&\multicolumn{4}{c}{methods}\\
    \cline{3-6}
    &&MI-FGSM & GRA & V-Cloak & proposed \\
    \hline
    \multicolumn{6}{c}{de-id}\\
\hline
ECAPA & w& 24.93& 46.44 &36.97 & 46.79 \\
    \hline
    ResNet34 &b & 7.99 & 12.15&9.82 &\textbf{16.21} \\
    \hline
    xi-vector &b & 15.37&23.32 &26.10 & \textbf{37.47}\\
    \hline
    \textcolor{black}{CAM++} & \textcolor{black}{b} &\textcolor{black}{ 12.67} & \textcolor{black}{15.37} & \textcolor{black}{\textbf{24.41}} &\textcolor{black}{ 23.72}\\
    \hline
    \textcolor{black}{ResNet100 }&\textcolor{black}{ b }&\textcolor{black}{ 4.01 }&\textcolor{black}{ 5.63 }& \textcolor{black}{2.48 }&\textcolor{black}{ \textbf{7.04}}\\
    \hline
    \multicolumn{6}{c}{id-unlnk}\\
\hline
ECAPA &w& 35.71 & 33.11 & 22.39 & 38.93 \\
    \hline
    ResNet34&b  & 11.90 & 13.20 & 11.36 & \textbf{19.47}\\
    \hline
    xi-vector &b & 21.74 & 24.06 & 17.81 & \textbf{31.19} \\
    \hline
    \textcolor{black}{CAM++} & \textcolor{black}{b} & \textcolor{black}{15.59} & \textcolor{black}{16.99} & \textcolor{black}{20.42} &\textcolor{black}{\textbf{ 23.70} }\\
    \hline
    \textcolor{black}{ResNet100} & \textcolor{black}{b} & \textcolor{black}{5.48} & \textcolor{black}{5.74} & \textcolor{black}{2.93} &\textcolor{black}{ \textbf{7.53} } \\
    \hline
    \end{tabular}
    \end{subtable}
\end{table}

The EERs obtained by the adversarial speech generation methods in both de-identification and identity unlinkability ASV evaluations are presented in Table \ref{tb. black-box ASV eval res}(\subref{tb: black-box asv res}). The MI-FGSM, GRA, V-Cloak, and our proposed methods are included. \textcolor{black}{From Table \ref{tb. black-box ASV eval res}(\subref{tb: black-box asv res}), it can be observed that, compared to the EERs obtained from the original recordings as presented in Table \ref{tb. black-box ASV eval res}(\subref{tb: black-box original}), the adversarial speech generated by the compared methods exhibited higher EERs when evaluated by the black-box extractors. This demonstrated their voice privacy protection effectiveness in black-box scenarios. Among the compared methods, in the de-identification evaluation, the proposed method obtained the second highest EER against the CAM++ speaker extractor and the highest EERs against the other three speaker extractors. In the identity unlinkability evaluations, the proposed method achieved the highest EERs against all four black-box extractors.} These results indicate the effectiveness of the proposed model in the transferability of voice privacy protection capabilities.

\begin{table}[t]
    \caption{\textcolor{black}{EERs(\%) obtained by the speaker-adversarial speech against the adaptive attack methods of median smoothing (MS), quantization (QT), low pass filtering (LPF), and AAC compression. Both the de-identification (de-id) and identity unlinkability (id-unlnk) performances are included. The results obtained by the MI-FGSM, GRA, V-Cloak, and the proposed method are presented. The performances obtained without attack are provided for reference in the row labeled \emph{w/o attack}. The average EERs obtained across all the attack methods are presented in the row \emph{Avg} for each compared method.}}
    \centering
    \begin{tabular}{c|c|c|c|c}
    \hline
    &\textcolor{black}{MI-FGSM} & \textcolor{black}{GRA} & \textcolor{black}{V-Cloak} &\textcolor{black}{ proposed }\\
    \hline
    \multicolumn{5}{c}{\textcolor{black}{de-id}}\\
    \hline
    \textcolor{black}{w/o attack} &\textcolor{black}{ 24.93}&\textcolor{black}{ 46.44} &\textcolor{black}{36.97} & \textcolor{black}{46.79} \\
    \hline
    \hline
    \textcolor{black}{MS}  & \textcolor{black}{16.55} &\textcolor{black}{ 30.23}&\textcolor{black}{12.08} &\textcolor{black}{\textbf{31.33}} \\
    \hline
   \textcolor{black}{ QT} &\textcolor{black}{ 5.49}&\textcolor{black}{\textbf{8.01}} &\textcolor{black}{4.17} &\textcolor{black}{ 7.63}\\
    \hline
   \textcolor{black}{ LPF} &\textcolor{black}{11.36}&\textcolor{black}{11.98} &\textcolor{black}{12.85} &\textcolor{black}{\textbf{13.59}}\\
    \hline
    \textcolor{black}{AAC} &\textcolor{black}{ 9.40}&\textcolor{black}{17.19} &\textcolor{black}{8.47} &\textcolor{black}{ \textbf{17.96}}\\
    \hline
    \hline
    \textcolor{black}{Avg} &\textcolor{black}{ 10.70} &\textcolor{black}{ 16.86} &\textcolor{black}{9.39} &\textcolor{black}{ \textbf{17.62}} \\
    \hline
    \multicolumn{5}{c}{\textcolor{black}{id-unlnk}}\\
    \hline
    \textcolor{black}{w/o attack} &\textcolor{black}{ 35.71} &\textcolor{black}{ 33.11} &\textcolor{black}{ 22.39} &\textcolor{black}{ 38.93} \\
    \hline
    \hline
   \textcolor{black}{ MS  }&\textcolor{black}{ 20.45} &\textcolor{black}{ 22.59} &\textcolor{black}{ 11.65} & \textcolor{black}{\textbf{29.42}}\\
    \hline
    \textcolor{black}{QT}  &\textcolor{black}{ 10.36} &\textcolor{black}{\textbf{ 12.21} } &\textcolor{black}{ 5.62} & \textcolor{black}{11.63 }\\
    \hline
   \textcolor{black}{ LPF} &\textcolor{black}{12.39} &\textcolor{black}{13.14} &\textcolor{black}{14.19}&\textcolor{black}{\textbf{15.66}}\\
    \hline
    \textcolor{black}{AAC}  &\textcolor{black}{ 12.23} &\textcolor{black}{ 14.86} &\textcolor{black}{ 10.33} & \textcolor{black}{\textbf{20.32}} \\
    \hline
    \hline
    \textcolor{black}{Avg} &\textcolor{black}{ 13.86} &\textcolor{black}{15.70} &\textcolor{black}{ 10.45} & \textcolor{black}{\textbf{19.26}} \\
    \hline
    \end{tabular}
    \label{tb: adaptive attack}
\end{table}

Compared with the white-box evaluations, the adversarial speech generated by the four methods exhibited decreased EERs in both the de-identification and identity unlinkability evaluations. This suggests a reduction in their efficacy of voice privacy protection capabilities against black-box speaker extractors. Such observations are consistent with the inherent black-box challenges in adversarial attack research \cite{zhu2023boosting}. Moreover, it is noteworthy that, in comparisons between the ResNet34 and xi-vector speaker extractors, which were trained with the same dataset, the xi-vector exhibited higher EERs than ResNet34 across all trials in both de-identification and identity unlinkability evaluations. This may be attributed to that the speaker encoder in our xi-vector model shared its architecture with that of the ECAPA-TDNN model, which was utilized to supervise our adversarial speech generation. \textcolor{black}{Furthermore, under the ResNet100 speaker extractor, which demonstrated strong speaker attribute extraction capabilities as evidenced by an EER of 0.58 on the original recordings, the voice privacy protection effectiveness of the compared methods significantly declined, ranging from approximately 2 to 8 in both evaluation scenarios. This indicates the vulnerability of the voice privacy protection methods to strong speaker extractors, warranting further investigation.}

\textcolor{black}{
\subsubsection{Voice privacy protection under adaptive attacks}
The voice privacy protection capabilities of MI-FGSM, GRA, V-Cloak, and the proposed method were evaluated and compared under adaptive attacks. Specifically, four attack methods that are commonly used to remove adversarial perturbations were examined, including median smoothing\cite{characterinzing} (kernel size of 3), quantization\cite{characterinzing} (quantization level $\lambda$ of $2^8$), low pass filtering \cite{chen2022towards} ($f_p=500$ Hz, $f_s=1000$ Hz), and AAC compression\cite{chen2022towards}. The experiments were conducted with the open-source code available at \footnote{\label{attack}\url{https://github.com/SpeakerGuard/SpeakerGuard}}. The results obtained on the VoxCeleb1-O test trials in both de-identification and identity unlinkability evaluations are presented in Table \ref{tb: adaptive attack}. \textcolor{black}{Among the attack methods, the quantization approach resulted in a marked deterioration of the EERs, indicating a degradation in voice protection efficacy. This should be due to our implementation of the quantization method following the configurations applied in \footref{attack}, which prioritized attack strength at the expense of speech quality, obtaining a PESQ value reduction to 3.43 in our proposed method.} Table \ref{tb: adaptive attack} shows that the proposed method achieved the highest EERs against the median smoothing, low pass filtering, and AAC compression attacks, and the second highest EERs against the quantization attack in both de-identification and identity unlinkability evaluations. Furthermore, it achieved the highest average EERs in both evaluations. These observations indicate the effectiveness of the proposed method against adaptive attacks. However, compared to the scenario where no attack was applied, the voice privacy protection capability of the proposed method decreased, highlighting the need for further enhancements in its resilience against adaptive attacks.}

\begin{table}[t]
\caption{Speech quality and voice privacy protection results obtained on the out-of-domain datasets, including the development and test subsets of LibriSpeech and AIShell, represented with Libri-dev, Libri-test, AIShell-dev, and AIShell-test, respectively. PESQ values and EERs (\%) in the de-identification (de-id) and identity unlinkability (id-unlnk) evaluations of the V-Cloak and proposed methods are presented. The results obtained on the in-domain test set VoxCeleb1-O are included for reference.}
\label{tb. out-of-domain asv res}
\centering
\begin{tabular}{c|c|c|c|c}
\hline
&\multirow{2}{*}{dataset}&speech qaulity&\multicolumn{2}{c}{voice privacy protection}\\
\cline{3-5}
&&PESQ&de-id&id-unlnk\\
\hline
\multirow{5}{*}{V-Cloak}&VoxCeleb1-O&3.22&36.97&22.39\\
&Libri-dev&3.06&31.48&16.65\\
&Libri-test&3.06&30.73&15.84\\
&AIShell-dev&2.97&38.48&21.54\\
&AIShell-test&2.96&42.17&17.84\\
\hline
\multirow{5}{*}{proposed}&VoxCeleb1-O&4.03&46.79&38.93\\
&Libri-dev&3.87&48.09&26.38\\
&Libri-test&3.88&45.42&30.27\\
&AIShell-dev&3.75&60.33&20.10\\
&AIShell-test&3.78&63.41&15.65\\
\hline
\end{tabular}
\end{table}

\subsubsection{Out-of-domain datasets}
The voice privacy protection capability of the speaker-adversarial speech generated with V-Cloak and the proposed method was evaluated on out-of-domain datasets, including the development and test subsets of LibriSpeech and AIShell. LibriSpeech is an English dataset while AIShell is in Chinese. From Table \ref{tb. out-of-domain asv res}, it can be noted that compared to the in-domain test set VoxCeleb1-O, the PESQ values decreased for both the LibriSpeech and AIShell datasets. Moreover, the adversarial speech produced by the proposed method demonstrated significantly higher PESQ values compared to those generated by the V-Cloak method on both the in-domain and out-of-domain datasets. This further demonstrates the superiority of the proposed method in preserving the quality of the original speech.

In the de-identification evaluations, the proposed method achieved EERs that were comparable to or even higher than VoxCeleb1-O, indicating its effectiveness in de-identification for the out-of-domain datasets. In comparison, the proposed method achieved higher EERs than the V-Cloak method in the de-identification evaluation for both the LibriSpeech and AIShell datasets, suggesting its superior performance in de-identification for out-of-domain datasets. In the identity unlinkability evaluation, both compared methods yielded lower EERs on the LibriSpeech and AIShell datasets than VoxCeleb1-O, indicating their reduced effectiveness in identity unlinkability on out-of-domain datasets. Furthermore, since English is a predominant language in the VoxCeleb1 and VoxCeleb2 training datasets, while Chinese is sparse, the English test dataset LibriSpeech did not outperform the Chinese test dataset AIShell. This indicates that language does not obviously affect the generalization ability of speaker-adversarial generation methods.

\subsubsection{\textcolor{black}{Stability}}
\textcolor{black}{Finally, as the proposed batch mean loss is stochastically implemented in every training process, the stability of the proposed method was examined. In this experiment, five additional models were trained independently and evaluated in terms of both de-identification and identity unlinkability. The results are provided in Table \ref{tb. stability eval res}. From the table, it can be observed that combined with the one examined in the previous evaluations, the six models achieved consistent performance in both evaluation scenarios. Moreover, their average performance was comparable to that of the model evaluated in previous experiments, further validating the effectiveness of the proposed method.}

\begin{table}[t]
\caption{\textcolor{black}{EERs(\%) obtained by the five models in both de-identification (de-id) and identity unlinkability (id-unlnk) evaluations. The model evaluated in the previous experiments is represented as ${\rm M}_0$, and the newly trained models are represented with ${\rm M}_1$ to ${\rm M}_5$, respectively. The average EERs obtained across the six models are presented in the column \emph{Avg}.}}
\label{tb. stability eval res}
\centering
\begin{tabular}{c|c|c|c|c|c|c|c}
\hline
& \textcolor{black}{${\rm M}_0$} &\textcolor{black}{ ${\rm M}_1$} &\textcolor{black}{ ${\rm M}_2$} &\textcolor{black}{ ${\rm M}_3$} &\textcolor{black}{ ${\rm M}_4$} &\textcolor{black}{ ${\rm M}_5$} &\textcolor{black}{Avg}  \\
\hline
\textcolor{black}{de-id} &\textcolor{black}{ 46.79}&\textcolor{black}{ 47.90} &\textcolor{black}{48.08}&\textcolor{black}{ 47.56} &\textcolor{black}{ 47.36} &\textcolor{black}{ 47.60}&\textcolor{black}{ 47.55}\\
\hline
\textcolor{black}{id-unlnk }&\textcolor{black}{ 38.93} &\textcolor{black}{ 39.00}&\textcolor{black}{ 38.86} &\textcolor{black}{ 38.84} &\textcolor{black}{ 38.85}&\textcolor{black}{38.95} &\textcolor{black}{ 38.91}\\
\hline
\end{tabular}
\end{table}

\section{Conclusions}
\label{sec. conclusion}
This paper focuses on the application of speaker-adversarial speech for asynchronous voice anonymization. Aimed at improving the identity unlinkability of anonymized utterances from the same speaker while avoiding the privacy violation of an actual speaker as present in the existing targeted training strategy, we propose an any-to-any training strategy. Specifically, a batch mean loss function is defined which supervises the anonymization of utterances from various original speakers within a training mini-batch to a common pseudo-speaker, approximated as the average speaker in the mini-batch. Furthermore, we introduce a speaker-adversarial speech generation model that integrates supervision from both the any-to-any and untargeted attack training settings. Experiments conducted on the VoxCeleb dataset demonstrated the efficacy of the proposed model in asynchronous voice anonymization in various aspects. Firstly, the proposed any-to-any strategy effectively enhanced identity unlinkability, thereby achieving a balance between the identity unlinkability and de-identification capabilities of anonymized speech. Secondly, the proposed method achieved enhanced de-identification and identity unlinkability than the optimization-based methods, including MI-FGSM and GRA, while obtaining comparable performance in preserving original speech quality, human perception, and intelligibility. Thirdly, our proposed generation model achieved superior performances to the existing voice-protection model, V-Cloak, in speech quality, human perception of speaker attribute, de-identification, and identity unlinkability.

Additionally, further experiments were conducted to examine the proposed method's voice protection efficacy against black-box speaker extractors \textcolor{black}{and adaptive attacks}, its generalization to out-of-domain datasets, \textcolor{black}{and stability}. Observations from the experimental results indicate that the speaker-adversarial speech exhibits a decline in voice privacy protection effectiveness when subjected to attacks by black-box speaker extractors. \textcolor{black}{Besides, the proposed method exhibited a degradation in voice privacy protection capability against adaptive attacks.} Moreover, the speaker-adversarial speech generation method demonstrated slightly reduced performance on out-of-domain datasets, especially in terms of identity unlinkability. 
Based on these observations, future research on the application of the proposed speaker-adversarial speech generation method in voice anonymization may be conducted on: 1) enhancing the transferability of protection capabilities against black-box speaker extractors, \textcolor{black}{2) improving the resilience against adaptive attacks, and 3)} enhancing the capability for identity unlinkability in out-of-domain test sets. 


\bibliographystyle{IEEEtran}



 

\begin{IEEEbiography}[{\includegraphics[width=1in,height=1.25in,clip,keepaspectratio]{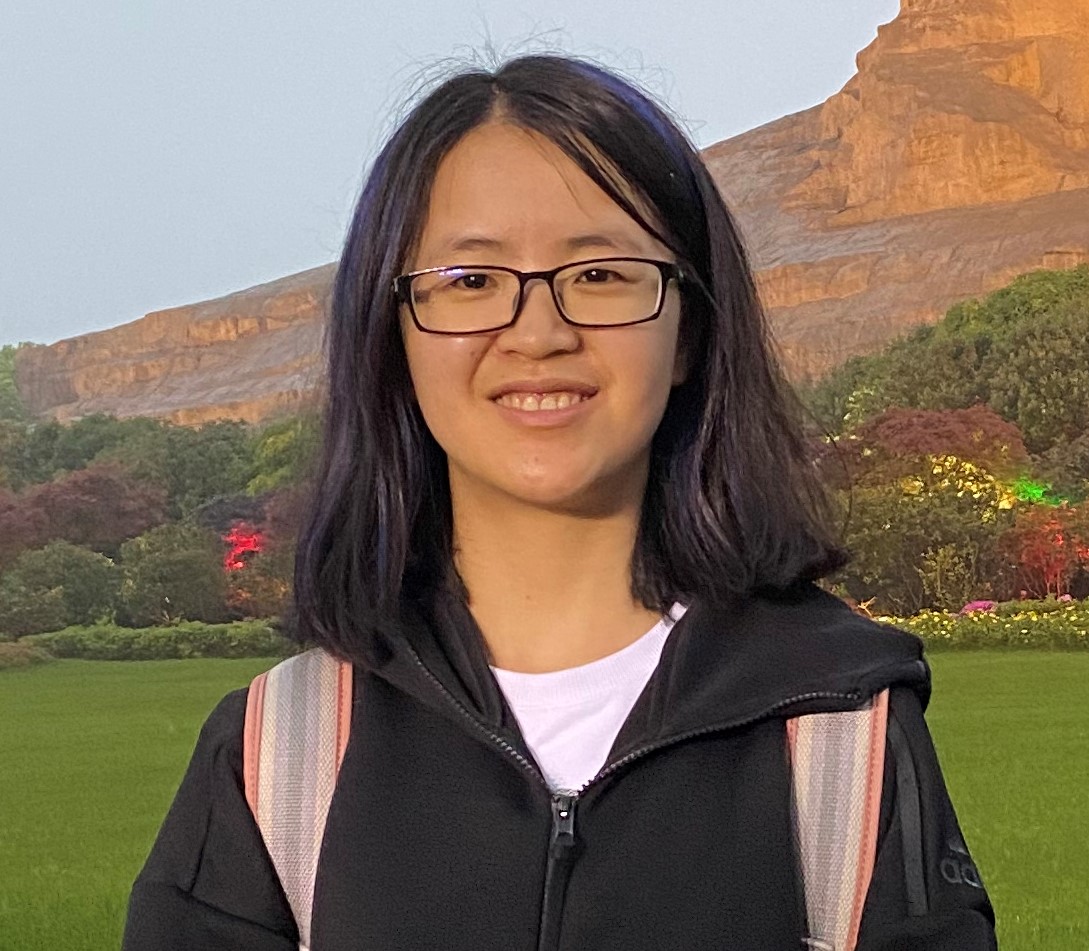}}]
{Liping Chen} (Senior Member, IEEE) received a Ph.D. degree in signal and information processing from the University of Science and Technology of China (USTC), Hefei, China, in 2016. From July 2016 to December 2022, she worked as a speech scientist at Microsoft. She is currently an Associate Researcher with the University of Science and Technology of China (USTC), Hefei, China. Her research interests include speech processing, voice privacy protection, speech synthesis, and speaker recognition.
\end{IEEEbiography}

\begin{IEEEbiography}
[{\includegraphics[width=1in,height=1.25in,clip,keepaspectratio]{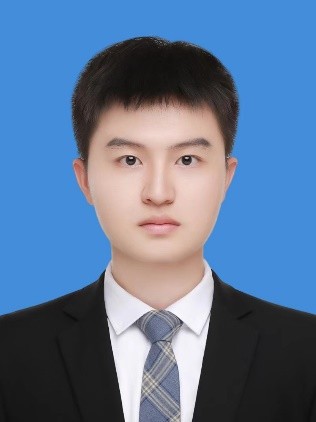}}]
{Chenyang Guo} is now a master student in signal and information processing from the University of Science and Technology of China (USTC), Hefei, China. His research interests include speech processing, voice privacy protection, and speaker-adversarial speech generation.
\end{IEEEbiography}

\begin{IEEEbiography}
[{\includegraphics[width=1in,height=1.25in,clip,keepaspectratio]{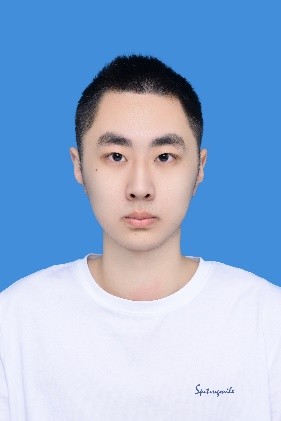}}]
{Rui Wang} is now a master student in signal and information processing from the University of Science and Technology of China (USTC), Hefei, China. His research interests include speech processing, voice privacy protection, and speech generation.
\end{IEEEbiography}

\begin{IEEEbiography}
[{\includegraphics[width=1in,height=1.25in,clip,keepaspectratio]{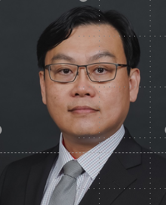}}]
{Kong Aik Lee} (Senior Member, IEEE) received the Ph.D. degree from Nanyang Technological University, Singapore, in 2006. From 2006 to 2018, he was a Research Scientist and then a Strategic Planning Manager (concurrent appointment) with the Institute for Infocomm Research, Singapore. From 2018 to 2020, he was a Senior Principal Researcher with the Data Science Research Laboratories, NEC Corporation, Tokyo, Japan. He was a Principal Scientist and Group Leader with the Agency for Science, Technology and Research (A*STAR), Singapore while holding a concurrent appointment as an Associate Professor with the Singapore Institute of Technology, Singapore. He is currently an Associate Professor with the Hong Kong Polytechnic University, Hong Kong. His research interests include the automatic analysis and privacy preservation of speaker characteristics, ranging from speaker recognition, language and accent recognition, voice biometrics, anti-spoofing, and voice anonymization. He was the recipient of the Singapore IES Prestigious Engineering Achievement Award 2013 and Outstanding Service Award by IEEE ICME 2020. Since 2016, he has been an Editorial Board Member of Elsevier Computer Speech and Language. From 2017 to 2021, he was an Associate Editor for IEEE/ACM TRANSACTIONS ON AUDIO,SPEECH, AND LANGUAGE PROCESSING. He is an elected Member of the IEEE Speech and Language Processing Technical Committee and was the General Chair of the Speaker Odyssey 2020 Workshop.
\end{IEEEbiography}

\begin{IEEEbiography}
[{\includegraphics[width=1in,height=1.25in,clip,keepaspectratio]{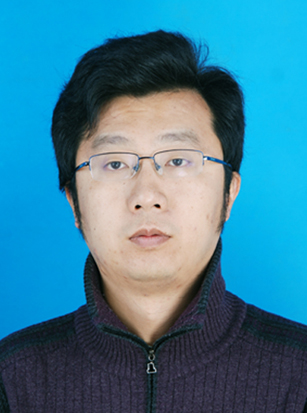}}]
{Zhen-Hua Ling} (M'10 SM'19) received a B.E. degree in electronic information engineering and M.S. and Ph.D. degrees in signal and information processing from the University of Science and Technology of China, Hefei, China, in 2002, 2005, and 2008. From October 2007 to March 2008, he was a Marie Curie Fellow with the Centre for Speech Technology Research, University of Edinburgh, Edinburgh, U.K. From July 2008 to February 2011, he was a joint Postdoctoral Researcher with the University of Science and Technology of China, and iFLYTEK Company, Ltd., Hefei, China. He also worked at the University of Washington, Seattle, WA, USA, as a Visiting Scholar from August 2012 to August 2013. He is currently a Professor with the University of Science and Technology of China. His research interests include speech processing, speech synthesis, voice conversion, and natural language processing. He was the recipient of the IEEE Signal Processing Society Young Author Best Paper Award in 2010. He was an Associate Editor of IEEE/ACM TRANSACTIONS ON AUDIO, SPEECH, AND LANGUAGE PROCESSING from 2014 to 2018.

\end{IEEEbiography}





\end{document}